\documentclass[doublecol]{epl2}
\usepackage{amssymb}
\usepackage{graphicx}
\usepackage{cite}
\newcommand{\tensor}[1]{\underline{#1}}

\begin{document}
\title{Ultrafast photoinduced phase separation dynamics in Pr$_{0.6}$Ca$_{0.4}$MnO$_3$ thin films}
\author{T. Mertelj\inst{1,2}, R. Yusupov\inst{1}, A. Gradi\v{s}ek\inst{1}, M. Filippi\inst{3} , W. Prellier\inst{3} and D. Mihailovic\inst{1,2}}
\date{\today}

\abstract{
The time resolved Magnetooptical Kerr effect in the substrate-strain-induced insulating ferromagnetic phase in (Pr$_{0.6}$Ca$_{0.4}$)MnO$_3$  thin films on two different substrates was measured in a magnetic field up to 1.1T. The photoinduced Kerr rotation and ellipticity show remarkably different magnetic-field dependence. From the comparison of the magnetic field dependencies of the  photoinduced and static Kerr rotation and ellipticity we conclude that a transient ferromagnetic metallic phase, embedded within the insulating ferromagnetic phase, is created upon the photoexcitation at low temperatures. A comparison of temporal dependence of the photoinduced Kerr signals with the photoinduced reflectivity indicates the change of the fractions of the phases takes place on a timescale of ten picoseconds independent of the substrate.
}

\institute{
    \inst{1}Jozef Stefan Institute, Jamova 39, 1000 Ljubljana, Slovenia, 
    
    \inst{2}Faculty of Mathematics and Physics, Univ. of Ljubljana, 1000 Ljubljana, Slovenia, 
    
    \inst{3}Laboratoire CRISMAT, CNRS UMR 6508, Bd du Marechal Juin, F-14050 Caen Cedex, France 
}

 {Magnetooptical effects}

\maketitle

\section{Introduction}

Photoinduced insulator-metal (IM) transition in (Pr,Ca)MnO$_3$\cite{MiyanoTanaka1997} and other related perovskite manganites is interesting not only due to its potential practical application, but also because of intimate connection between the metallic state and ferromagnetism in the manganites\cite{SantenJonker1950,SchifferRamirez1995,TomiokaAsamitsu1996}. Experiments indicate that a photoexcited metallic state forms on a sub-ps scale\cite{FiebigMiyano2000,KidaTonouchi2001} and that the ultrafast transition to the metallic state can be triggered by a direct excitation of the lattice degrees of freedom\cite{RiniTobey2007}. 

It is not clear, however, whether  on the sub-ps timescale the transient photoinduced metallic phase is the same as the static ferromagnetic metallic percolating phase.\cite{UeharaMori1999} There are some indications from time resolved magnetooptical Kerr effect (TRMOKE) measurements that the short range ferromagnetic order might be created simultaneously with the metallic state on the sub-ps timescale.\cite{MiyasakaNakamura2006,MatsubaraOkimoto2007} However, the difficulty is that in the manganites the TRMOKE response might not directly reflect magnetization dynamics\cite{McGillMiller2004} as in the transition-metal ferromagnets\cite{KoopmansvanKampen2000,BigotGuidoni2004} due to relatively slow charge dynamics\cite{LobadTaylor2000,MerteljBosak2000} and/or presence of phase separation\cite{CoxRadaelli1998,UeharaMori1999,LiuMoritomo2001,SaurelBrulet2006}, which both can result in nonmagnetic contributions to TRMOKE.

The fundamental questions how and on which timescale the transient metallic phase transforms into the metastable ferromagnetic metallic phase are therefore still open. The local ferromagnetic order can develop simultaneously with the metallic phase on a sub-ps timescale followed by slower reorientation of the local magnetization towards effective magnetic field. Alternatively, the transient metallic phase can be nonmagnetic on the sub-ps timescale nad the metastable ferromagnetic phase grows in form of clusters on a slower timescale.

In order to answer these questions we investigate the photoinduced-phase formation dynamics in the strain-induced ferromagnetic insulating state of Pr$_{0.6}$Ca$_{0.4}$MnO$_3$ thin films\cite{NelsonHill2004,MerteljYusupovAPL2008} by means of ultrafast TRMOKE. By comparing the photoinduced reflectivity with TRMOKE  and analyzing magnetic field dependence of both responses in addition to their time evolution at two different photon energies we are able to disentangle magnetic and nonmagnetic contributions to TRMOKE and detect \emph{a transient modulation of the individual fractions of coexisting phases} which appears on a timescale of ten picoseconds. We find no clear evidence that the magnetization is modified on the sub-picosecond scale. This suggests that the conducting ferromagnetic phase similar to the static one is not formed immediately on the sub-ps timescale.

\section{Experimental}
Thin films of Pr$_{0.6}$Ca$_{0.4}$MnO$_3$ (PCMO60) with the thickness of $\sim$3000\AA \ were grown on (001)-oriented
SrTiO$_3$ (STO) and LaAlO$_3$ (LAO) substrates as described elsewhere.\cite{PrellierSimon2000} STO induces tensile and LAO compressive strain in the film.\cite{NelsonHill2004}  Due to the strain PCMO60 thin films are ferromagnetic and insulating below $\sim$120K.\cite{NelsonHill2004,MerteljYusupovAPL2008} In PCMO60/STO static magnetization and MOKE measurements indicate that at 5K a metastable static ferromagnetic metallic (SFM) phase coexists with a stable ferromagnetic insulating (FI) phase  at the surface of the film in a magnetic field below 1.1 T already. Near the surface, within the penetration depth of our optical probe, which is $\sim$50 nm, both films show out-of-plane hard-axis magnetic anisotropy of the FI phase.  In PCMO60/STO the SFM phase also shows out-of-plane hard-axis magnetic anisotropy with a hysteresis in the out-of-plane magnetic field which presumably originates from Bloch domain walls, as described in detail in ref. \cite{MerteljYusupovAPL2008}.

A linearly polarized pump beam with the photon energy 1.55 eV, the pulse length 50 fs and repetition frequency 250 kHz was focused to a 250-$\mu$m diameter spot on the sample in a nearly perpendicular geometry. The fluence of the pump pulses was 150$\mu$J/cm$^2$. The weaker probe beam with the photon energy $\hbar \omega_{probe}=$  1.55 eV or 3.1 eV and the diameter 220 $\mu$m was focused to the same spot with the polarization perpendicular to that of the pump. The reflected probe beam was analyzed by a Wollaston prism and a pair of balanced silicon PIN photodiodes using standard lockin techniques.

Samples were mounted on a cold finger of an optical liquid-He flow cryostat equipped with CaF$_2$ windows placed in an 1.1-T electromagnet with hollow poles. All MOKE measurements were conducted in the polar geometry with the magnetic field perpendicular to the film. During TRMOKE measurements the static Kerr rotation $\phi_\mathrm{K} = Re\left(\Theta _\mathrm{K}\right)$ was compensated by a computer controlled rotation stage to keep the detector balanced at any magnetic field value.
The photoinduced complex Kerr angle transients $\Delta \Theta _\mathrm{K}=\frac{1}{2}\left[ \Delta \Theta \left( H\right) -\Delta \Theta \left( -H\right) \right] $  were obtained by subtracting the photoinduced probe polarization change $\Delta \Theta$ taken at two opposite directions of the magnetic field to remove nonmagnetic contributions.
For all optical measurements the samples were first zero-field cooled (ZFC) to the lowest temperature. Subsequently data were collected at fixed temperatures during the warming part of a cycle.

\begin{figure}[h]
  \begin{center}
  \includegraphics[angle=-90,width=0.47\textwidth]{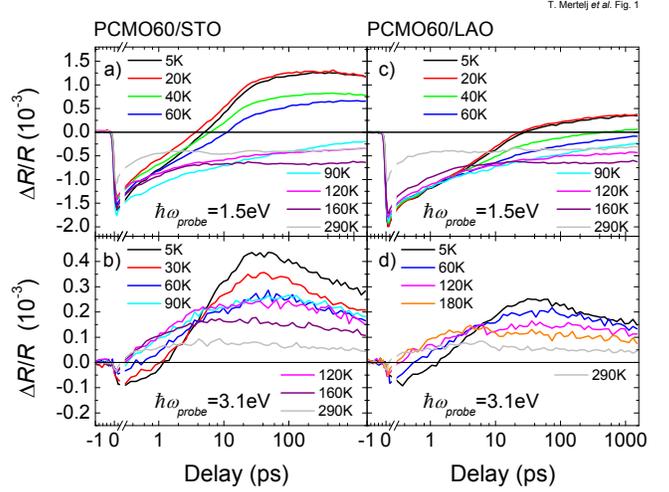}
  \end{center}
  \caption{Temperature dependence of  reflectivity transients in zero magnetic field at 1.55-eV probe-photon energy for the PCMO60/STO and the PCMO60/LAO  sample a) and c) respectively. For comparison reflectivity transients at 3.1-eV probe-photon energy are shown in b) and d).}
\end{figure}

\section{Photoinduced reflectivity}
In Fig. 1 we show the temperature dependence of reflectivity transients for two probe-photon energies in the absence of magnetic field. At the room temperature a negative sub-picosecond transient is followed by a plateau independently on the probe-photon energy (PPE). At $\sim$120K, concurrent with the onset of ferromagnetic ordering, another relaxation component appears with a risetime on a 10-ps timescale. This component decays on a timescale of a few hundred ps at 3.1-eV PPE while at 1.55-eV a decay becomes apparent only for the delay beyond a ns.

The transients display no anisotropy when the polarizations are rotated with respect to the sample. At 3.1-eV PPE the transients show no magnetic field dependence, while at 1.55-eV PPE a small reversible increase of magnitude of the 10-ps component was observed in both samples. In the PCMO60/STO sample we observed in addition to the reversible also an irreversible increase during the first application of the magnetic field to the ZFC sample. All optical data on the PCMO60/STO sample reported here were therefore taken after the sample was switched into the state with no observable irreversibility.

\section{TRMOKE: magnetic field dependence}
In contrast to the time dependence of the photoinduced reflectivity the magnetic field dependence of the MOKE response strongly depends on the substrate and PPE\cite{MerteljYusupovAPL2008}. Let us discuss magnetic field dependence of TRMOKE at nanosecond delays first.

\begin{figure}[h]
  \begin{center}
  \includegraphics[angle=-90,width=0.47\textwidth]{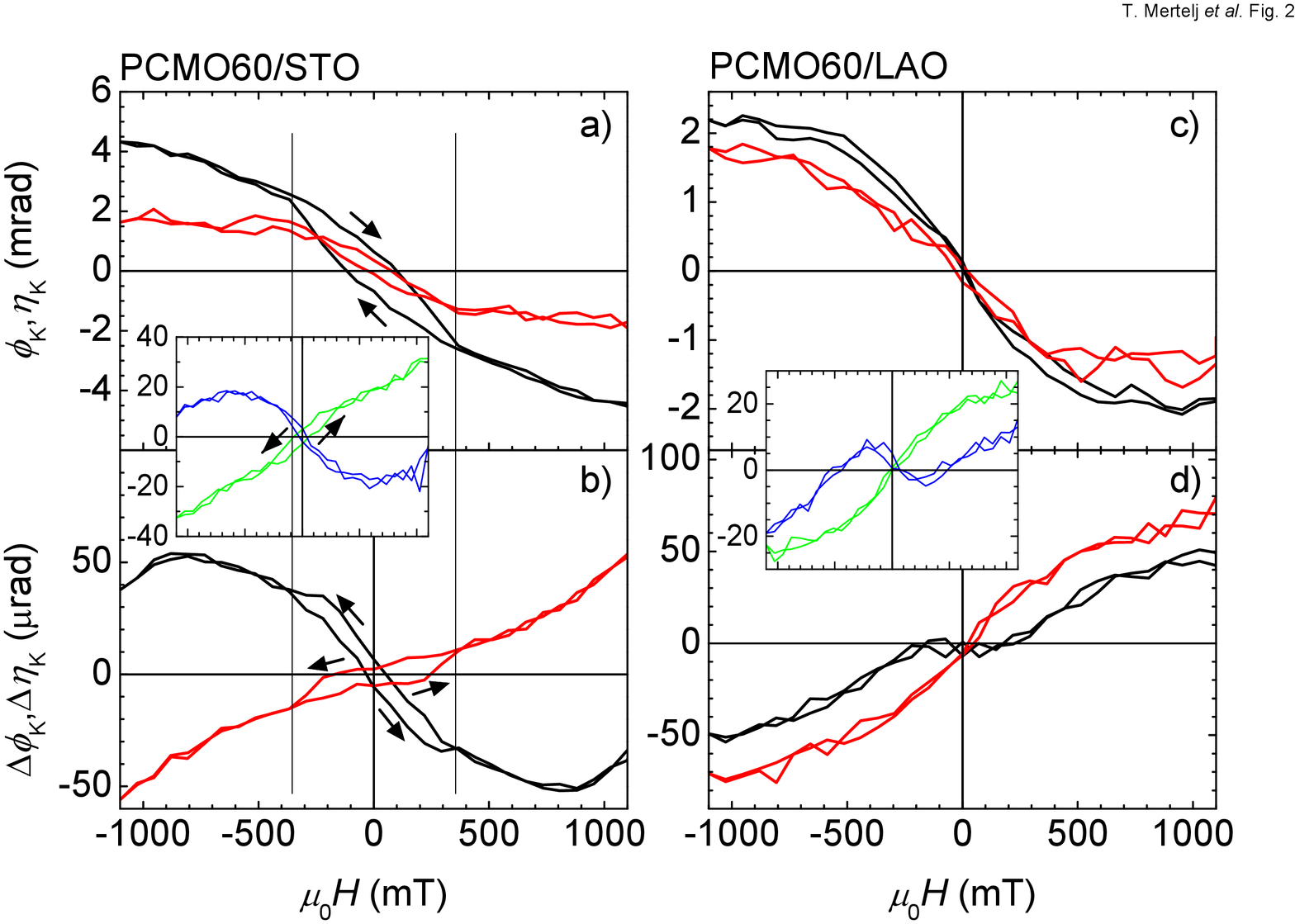}
  \end{center}
\caption{Magnetic field dependence of the static Kerr rotation,  $\phi_\mathrm{K}$, (black/dark) and ellipticity, $\eta_\mathrm{K}$, (red/light) at 5K and 3.1eV probe-photon energy for the PCMO60/STO sample a) and the PCMO60/LAO sample c). The corresponding photoinduced counterparts $\Delta\phi_\mathrm{K}$ (black/dark) and  $\Delta\eta_\mathrm{K}$ (red/light)  at 1500 ps  are shown in b) and d). $\Delta\phi_\mathrm{K}$ (green/light) and  $\Delta\eta_\mathrm{K}$ (blue/dark) at 1.55eV probe-photon energy are shown in the insets.}
\label{fig:H-loops}
\end{figure}

In the PCMO60/STO sample a clear hysteresis is present in the static Kerr rotation, $\phi _\mathrm{K} = Re(\Theta _\mathrm{K})$, at 3.1-eV PPE while the static Kerr ellipticity, $\eta _\mathrm{K} = Im(\Theta _\mathrm{K})$, displays no hysteresis at both PPE (see Fig. 2(a)) indicating the presence of the FI and SFM phases\cite{MerteljYusupovAPL2008}. A clear hysteresis is also present in most of the components of the photoinduced complex Kerr angle except in  the photoinduced Kerr ellipticity, $\Delta\eta _\mathrm{K}$, at 1.55-eV PPE (see inset to Fig. 2b). The magnetic field dependence of $\Delta\phi_\mathrm{K}$ at 3.1-eV PPE, when ignoring the hysteresis, is very similar to $\Delta\eta _\mathrm{K}$ at 1.55-eV PPE. Simultaneously, the magnetic field dependence of $\Delta\eta _\mathrm{K}$ at 3.1-eV PPE is similar to $\Delta\phi _\mathrm{K}$ at 1.55-eV PPE.
All are different from their static counterparts, $\phi _\mathrm{K}$ and $\eta _\mathrm{K}$.

Comparison of $\Delta\phi _\mathrm{K}$ with $\phi _\mathrm{K}$ at 3.1-eV PPE also reveals different relative sign of hysteresis loops with respect to the non-hysteretic background. While the non-hysteretic background in $\Delta\phi _\mathrm{K}$ has the same sign as the background in $\phi _\mathrm{K}$, the loops display opposite sense of rotation indicating a negative sign.

Contrary to the PCMO60/STO sample there is no evidence for the presence of the SFM phase in addition to the FI phase below 1.1 T in  the PCMO60/LAO sample in $\phi _\mathrm{K}$ and $\eta _\mathrm{K}$\cite{MerteljYusupovAPL2008}, which have the same magnetic field dependence with no hysteresis.  This is, however, \emph{not the case} for the photoinduced Kerr rotation, $\Delta\phi _\mathrm{K}$, at 3.1-eV PPE and ellipticity, $\Delta\eta _\mathrm{K}$, at 1.55-eV PPE. They both display kinks at $\sim\pm200$ mT which are not present in the static counterparts nor in the SQUID magnetization loops\cite{MerteljYusupovAPL2008}. In the region of the kinks the photoinduced change is positive with respect to the static value. Qualitatively the behavior is similar to PCMO60/STO where kinks extend over a broader field range.
The kinks are not compatible with the static FI magnetic phase (nor with the SFM phase in PCMO60/STO sample) and indicate the presence of another photoinduced ferromagnetic phase which we associate with the photoinduced transient ferromagnetic metallic (PTFM) phase.

\section{TRMOKE: time dependence}

\begin{figure}[h]
  \begin{center}
    \includegraphics[angle=-90,width=0.47\textwidth]{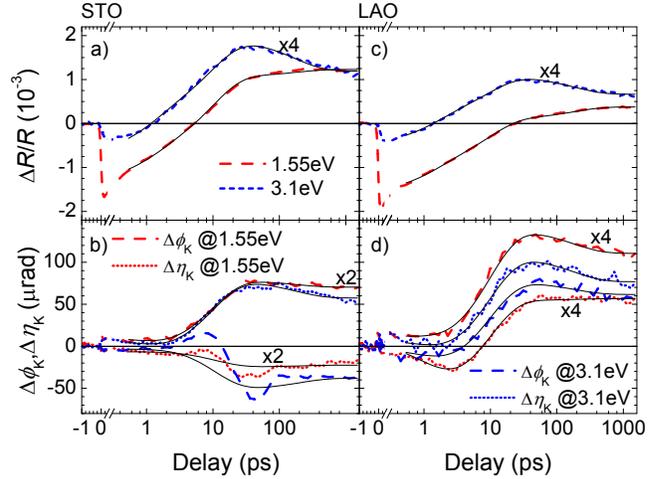}
  \end{center}
  \caption{Comparison of the photoinduced reflectivity ($\mu_0 H=0$ T) a), c) and the complex Kerr angle ($\mu_0H=\pm$1.1 T) b), d) at 5K and different probe-photon energy for the PCMO60/STO sample a), b) and the PCMO60/LAO sample c), d).  The thin solid lines represent fits discussed in text.} 
\label{fig:TRMOKE-dly}
\end{figure}

\begin{figure}[h]
  \begin{center}
  \includegraphics[angle=-0,width=0.40\textwidth]{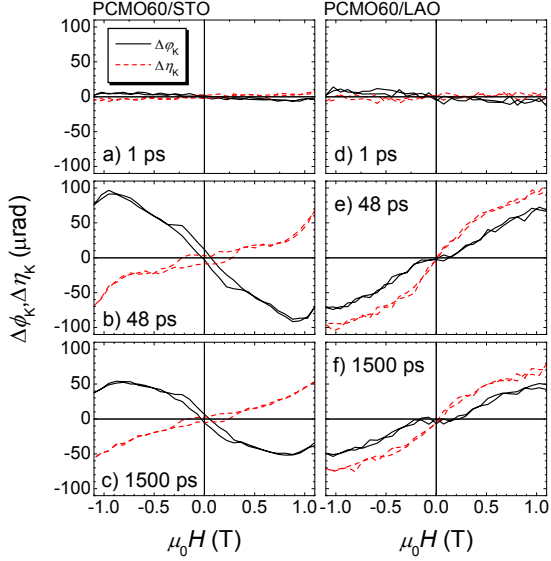}
  \end{center}
  \caption{TRMOKE $H$-loops at 3.1-eV PPE energy and 5K at representative delays.}
\label{fig:dly-evol-both-400}
\end{figure}

Let us now discuss the time dependence of TRMOKE shown in Fig. \ref{fig:TRMOKE-dly}b), \ref{fig:TRMOKE-dly}d) and Fig. \ref{fig:dly-evol-both-400}. Similar to the magnetic field dependence the delay dependence of $\Delta\eta _\mathrm{K}$ at 1.55-eV PPE shows similar behavior as $\Delta\phi _\mathrm{K}$ at 3.1-eV PPE and the delay dependence of $\Delta\phi _\mathrm{K}$ at 1.55-eV PPE shows similar behavior as $\Delta\eta _\mathrm{K}$ at 3.1-eV PPE (see Fig. \ref{fig:TRMOKE-dly})  so it is enough to discuss the traces at 3.1-eV PPE only.

On a sub-ps timescale we observe a resolution limited onset of both  $\Delta\phi _\mathrm{K}$ and  $\Delta\eta _\mathrm{K}$ followed by a weakly time dependent plateau extending to $\sim$2 ps. The plateau has a different sign for $\Delta\phi _\mathrm{K}$ and  $\Delta\eta _\mathrm{K}$. In the plateau region no hysteresis is observed in $H$-loops of the photoinduced signal (see Fig \ref{fig:dly-evol-both-400} a), d)). Following the plateau $\Delta\eta _\mathrm{K}$ rises on a 10-ps timescale and then partially relaxes to a finite value on a few-hundred-ps timescale. $\Delta\phi _\mathrm{K}$ behaves similarly to $\Delta\eta _\mathrm{K}$ in the PCMO60/LAO sample only. In the PCMO60/STO sample however, there is an additional strongly-damped oscillatory component present in $\Delta\phi _\mathrm{K}$  on top of the 10-ps component and the sign of the 10-ps component is negative. There is no change of the shape of $H$-loops with  delay after initial rise in the PCMO60/LAO sample (see Fig \ref{fig:dly-evol-both-400} e), f)). In PCMO60/STO some change in the shape of $H$-loops is observed in the region where the oscillatory component is present (see Fig. \ref{fig:dly-evol-both-400} b), c)).

Comparison of the reflectivity transients with TRMOKE transients in Fig. \ref{fig:TRMOKE-dly} indicates that with exception of the sub-ps dynamics the same relaxation components are present in all the traces of a given sample at a given temperature. Moreover, for the delay longer than 0.5 ps it is possible to fit all the traces of a given sample with a sum of three exponential functions: $y(t)=A_{0}+\sum A_{i}\exp \left( -t/\tau _{i}\right)$, where the relaxation times (see Table \ref{tab:tau}) \emph{are shared among all the traces} and the amplitudes $A_0$ and $A_i$ are independent parameters for each trace separately. The only exceptions are $\Delta\eta _\mathrm{K}$ at 1.55-eV PPE and $\Delta\phi _\mathrm{K}$ at 3.1-eV PPE in the PCMO60/STO sample where the strongly-damped oscillatory component is observed. While the oscillatory component shows magnetic field dependence there is virtually no magnetic field dependence of the relaxation times.

\begin{table}[t]
\centering
\begin{tabular}{l|c|c|c}
 & $\tau_1$ (ps) & $\tau_2$ (ps) & $\tau_3$ (ps)\\
\hline
PCMO60/LAO &1.3 $\pm$  0.1 & 9.8 $\pm$  0.2 & 182 $\pm$  11 \\
PCMO60/STO &1.1 $\pm$  0.1 & 9.9 $\pm$  0.3 & 177 $\pm$  16

\end{tabular}
\caption{The relaxation times obtained from the fits. For details see text.}
\label{tab:tau}
\end{table}

\section{Discussion}

In the case of a single magnetic component the photoinduced change of the complex Kerr angle in the polar configuration is given to the lowest order in magnetization,  

\begin{eqnarray}
\Delta \phi_\mathrm{K}(H_z,t)&=&\Delta M_z(t) g_r  + M_z(H_z) \Delta g_r(t) , \label{eq:Dphik}\\
\Delta \eta_\mathrm{K}(H_z,t)&=&\Delta M_z(t) g_i  + M_z(H_z) \Delta g_i(t) , \label{eq:Detak}
\end{eqnarray} 
where $g_r$ and $g_i$ represent magnetooptical couplings and $M_z$ the component of the magnetization perpendicular to the sample surface. The photoinduced change in the magnetooptical couplings results in contributions to  $\Delta \phi_\mathrm{K}(H_z,t)$ and $\Delta \eta_\mathrm{K}(H_z,t)$  which have the same shape of $H$-loops (including the hysteresis) as $M_z(H_z)$. The same holds when only the magnitude of the magnetization is changed ($\Delta M_z(t)\propto{M_z(H_z)}$), i.e. in the case of demagnetization or change of the volume fraction of a preexisting magnetic phase. Any other changes in $\Delta M_z(t)$ result in contributions to $\Delta \phi_\mathrm{K}(H_z,t)$ and $\Delta \eta_\mathrm{K}(H_z,t)$ which in general have different shape of the $H$-loops than $M_z(H_z)$, however, have \emph{identical shape for both}  $\Delta \phi_\mathrm{K}(H_z,t)$ and $\Delta \eta_\mathrm{K}(H_z,t)$.

In our experiment neither of these simple cases is observed. It is obvious that in both samples $H$-loops of $\Delta \phi_\mathrm{K}$ and $\Delta \eta_\mathrm{K}$ at any PPE can not be obtained by a linear combination of the static $\phi_\mathrm{K}$ and $\eta_\mathrm{K}$. This rules out a simple demagnetization and/or change of the magnetooptical couplings (due to the charge dynamics) of two independent preexisting magnetic phases as the source of the TRMOKE response. 

There are two more indications that the charge dynamics can not be the main origin of the TRMOKE dynamics. First, it is very unlikely that  the charge dynamics of different phases would have the same delay dependence. Second,  $\Delta\phi _\mathrm{K}/\phi _\mathrm{K}$ and  $\Delta\eta _\mathrm{K}/\eta _\mathrm{K}$ are ten times larger than $\Delta R/R$ indicating that the effect is not due to the photoinduced modification of optical-dipole-transition oscillator strengths which would be a result of the excited states absorption.

On the other hand, the TRMOKE dynamics also can not be governed directly by $\Delta M_z(t)$ because $\Delta \phi_\mathrm{K}$ and $\Delta \eta_\mathrm{K}$ $H$-loops differ and because the same relaxation components are observed in the photoinduced reflectivity transients in the absence of magnetic field and TRMOKE transients in a finite magnetic field. 

In a multiphase system there is another possible mechanism contributing to the TRMOKE and photoinduced reflectivity dynamics.  In a mixture of phases the effective dielectric function $\tensor{\epsilon}(\omega)$ is a function of the fraction $c$ and the dielectric functions $\tensor{\epsilon}_i(\omega)$ of different phases. If $c$ is time dependent $\Delta c(t)$ would be present in optical transients corresponding to the symmetrical part as well as the antisymmetrical part of $\tensor{\epsilon}(\omega)$ in addition to any internal dynamics of the involved phases. Since in our case the delay dependencies of all signals clearly contain some common components we propose that the common components originate from \emph{the dynamical phase transition} between different coexisting phases. The most obvious common component which we associate with $\Delta c(t)$ is the 10-ps component which consistently dominates in all the signals in both samples. 

\begin{figure}[h]
  \begin{center}
  \includegraphics[angle=-90,width=0.47\textwidth]{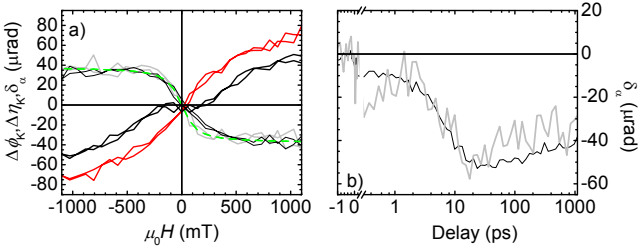}
  \end{center}
\caption{a) The linear combinations $\delta_{3.1}$ (thick grey line) and $\delta_{1.55}$ (thin black line). The the dashed green line is a Brillouin function fit to $\delta_{3.1}$. For comparison  $\Delta\phi_\mathrm{K}$ (black/dark) and  $\Delta\eta_\mathrm{K}$ (red/light)  at 1500 ps are also shown. b) The delay dependence of $\delta_{3.1}$ (thick grey line) and $\delta_{1.55}$ (thin black line). } \label{fig:PCMO60-LAO-diff}
\end{figure}

This is the most clearly seen in the PCMO60/LAO sample where a single static magnetic (FI) phase is present in our range of applied magnetic field \cite{MerteljYusupovAPL2008} in absence of the photoexcitation. It is obvious that $H$-loops of $\Delta\phi_\mathrm{K}$ ($\Delta\eta_\mathrm{K}$) (see Fig. \ref{fig:H-loops} d)) can be decomposed into two main components. The first which is proportional to the static $\phi_\mathrm{K}$ and $\eta_\mathrm{K}$ at 3.1-eV PPE is present with a negative sign (with respect to the static $\phi_\mathrm{K}$ and $\eta_\mathrm{K}$)  in both $\Delta\phi_\mathrm{K}$ and $\Delta\eta_\mathrm{K}$ and represents the transient decrease of the amount of the FI phase. The second, which corresponds to the kinks at $\sim\pm200$ mT, is extracted by forming linear combinations $\delta_{3.1}=\Delta\phi_\mathrm{K}-1.2 \Delta\eta_\mathrm{K}$ at 3.1-eV PPE and $\delta_{1.55}=\Delta\eta_\mathrm{K}-2 \Delta\phi_\mathrm{K}$ at 1.5-eV PPE (see Fig Fig. \ref{fig:PCMO60-LAO-diff}a)). We assign it to the photoinduced transient ferromagnetic metallic (PTFM) phase. 

From a fit of the Brillouin function to $\delta_{3.1}$ and $\delta_{1.55}$ we can infer that the PTFM phase is formed in form of clusters which consist of $\sim$500 spins. The delay dependencies of $\delta_{3.1}$ and $\delta_{1.55}$ beyond 1 ps (see Fig. \ref{fig:PCMO60-LAO-diff} b)) are both  virtually the same as the delay dependencies of $\Delta\phi_\mathrm{K}$ and $\Delta\eta_\mathrm{K}$ confirming the assignment of the 10-ps component to $\Delta c(t)$.

Since $\Delta\eta_\mathrm{K}$ at 3.1-eV PPE does not contain a significant contribution from the PTFM phase we use the ratio $\Delta\eta_\mathrm{K}/\eta_\mathrm{K}$ to estimate the transient decrease of the amount of the FI phase to be $\sim$3\%. This fraction does not necessarily correspond to the total amount of the PTFM phase because it is not necessary that all of the destroyed FI phase is converted to the PTFM phase.

In the case of the PCMO60/STO sample a unique decomposition of the $\Delta\phi_\mathrm{K}$ ($\Delta\eta_\mathrm{K}$) $H$-loops into components does not seem possible, presumably due to the presence of more than two components. In contrast to the PCMO60/LAO sample there is an additional SFM phase, responsible for the hysteretic component of the static $\phi_\mathrm{K}$ at 3.1-eV PPE \cite{MerteljYusupovAPL2008}, which coexists with the FI phase even in the absence of photoexcitation. The positive $\Delta\phi_\mathrm{K}/\phi_\mathrm{K}$ at 3.1-eV PPE and $\Delta\eta_\mathrm{K}/\eta_\mathrm{K}$ at 1.55-eV PPE suggest however, similarly as in the PCMO60/LAO sample, a transient emergence of a PTFM phase upon photoexcitation. This phase can be either created separately from the SFM phase or by growth of the preexisting SFM clusters. The opposite sense of rotation of the hysteretic part of $\Delta\phi_\mathrm{K}$ with respect to the static $\phi_\mathrm{K}$ at 3.1-eV PPE suggests demagnetization and/or decrease of the amount of the SFM phase. The TFM phase is therefore most likely created \emph{independently} of the preexisting SFM phase in form of clusters embedded within the FI phase. The size of the clusters appears smaller than in the PCMO60/STO sample judging from the absence of any apparent saturation in $\Delta\phi_\mathrm{K}$  below $\sim$500 mT. While the magnitude of $\Delta\eta_\mathrm{K}/\eta_\mathrm{K}$ at 3.1-eV PPE indicates that the amount of the destroyed FI phase is similar to the PCMO60/LAO sample, the magnitudes of $\Delta R/R$ and the maximum of $\Delta\phi_\mathrm{K}/\phi_\mathrm{K}$ suggest that the amount of the PTFM phase is larger in the PCMO60/STO sample.

After assignment of the 10-ps component to the risetime of $\Delta c(t)$ we turn now to the analysis of the dynamics on slower timescales. The slowest 200-ps component has different signs with respect to the 10-ps component when comparing $\Delta R/R$ at different PPE so it can not have the same origin as the 10-ps component and does not correspond the decay of the PTFM phase back to the FI phase. It is therefore attributed to the internal dynamics of the individual phases. Since the clusters of the PTFM phase show no clear decay in our data up to the longest 1.5-ns delay and in our experiment a new pump pulse arrives each 4 $\mu$s it is possible that the PTFM clusters do not decay completely before arrival of the next pump pulse.

On the sub picosecond timescale, $\Delta\phi _\mathrm{K}$ and  $\Delta\eta _\mathrm{K}$ have inconsistent signs with respect to the 10-ps component when we compare both samples so they can not correspond to $\Delta c(t)$ as well. Due to different signs of $\Delta\phi _\mathrm{K}/\phi _\mathrm{K}$ and  $\Delta\eta _\mathrm{K}/\eta _\mathrm{K}$  it is also very unlikely that they represent the sub-ps magnetization change.  We believe that they rather originate in the charge dynamics corresponding to electron-electron and electron-phonon interactions.\cite{LobadTaylor2000}  

Finally, the strongly-damped oscillatory component observed only in the PCMO60/STO sample is compatible with almost critically damped magnetization precession. Since it appears in the same TRMOKE components as the signature of the PTFM phase it could be associated with the magnetization reorientation of the PTFM-phase clusters.

It is important to note that the influence of the different substrate-induced strain is mainly reflected in experimental features associated with magnetization reorientation such as: hysteresis, the shape of $H$-loops and precession, and not in the phase-fraction dynamics which might be a more general feature. Indeed, photoinduced reflectivity transients similar to the 10-ps component were observed also in the ferromagnetic metallic manganites. The transients display increasing relaxation time and amplitude near the temperature induced IM  transition\cite{LobadTaylor2000,LiuMoritomo2000,MerteljMihailovic2003} and were associated with the demagnetization due to the spin-lattice relaxation\cite{LobadTaylor2000}. This is in contradiction with recent TRMOKE results\cite{McGillMiller2005} which indicate a possibility of a transient remagnetization in addition to demagnetization near the Curie temperature. We believe that this is a strong indication that the change of the fraction of the coexisting phases due to \emph{the photoinduced dynamical phase transition} contributes also to the slow photoinduced reflectivity transients observed near the IM transition in the ferromagnetic metallic manganites.

\section{Conclusions}
We measured magnetic field dependence of the time-resolved  photoinduced reflectivity and TRMOKE in (Pr$_{0.6}$Ca$_{0.4}$)MnO$_3$  thin films on two different substrates with different substrate-induced strain. By comparing the time evolutions of the photoinduced reflectivity with TRMOKE at two different probe photon energies we are able to separate different contributions to the TRMOKE dynamics. We find no clear evidence that the magnetization is modified on the sub-picosecond scale which leads us to believe that the conducting ferromagnetic phase similar to the static one is not formed immediately on the sub-ps timescale. Our data indicate the \emph{the photoinduced dynamical phase transition} of the FI phase into a photoinduced transient FM phase occurs on a slower 10-ps timescale independently of the substrate. The photoinduced transient FM phase is formed in clusters, which are $\sim$ 500 spins large in PCMO60/LAO and are smaller in PCMO60/STO. The photoinduced transient FM phase shows no clear decay within our time window of 1.5 ns. 

On the basis of our results we propose that the photoinduced dynamical phase transition is a universal feature of the phase separated state in manganites and contributes also to the slow photoinduced reflectivity transients observed near the IM transition\cite{LobadTaylor2000,LiuMoritomo2000,MerteljMihailovic2003} in the large-bandwidth ferromagnetic metallic manganites.

\begin{acknowledgments}
We acknowledge fruitful discussions with V.V. Kabanov. The work was supported within the FP6, project NMP4-CT-2005-517039
(CoMePhS).

\end{acknowledgments}

\bibliographystyle{eplbib}
\bibliography{biblio}

\end{document}